\documentclass[aps,pra,twocolumn,a4paper,showpacs,floatfix,
superscriptaddress, showpacs, preprintnumbers]{revtex4-1}

\usepackage{graphicx}
\usepackage{calc}
\usepackage{amsmath}
\usepackage{amssymb}
\usepackage{xspace}

\newcommand{\ud}{\mbox{\rm d}}
\newcommand{\Tr}{\mbox{\rm Tr}}
\newcommand{\I}{\mbox{\rm I}}

\newcommand{\ket}[1]{\left|\mathrm{#1}\right\rangle}

\newcommand{\D}{\mbox{\rm d}}

\newcommand{\AppOIOR}{Appendix~A\xspace}
\newcommand{\AppQSIOR}{Appendix~B\xspace}
\newcommand{\AppGB}{Appendix~C\xspace}
\newcommand{\AppWeber}{Appendix~D\xspace}
\newcommand{\AppPDTC}{Appendix~E\xspace}
\newcommand{\AppBell}{Appendix~F\xspace}
\newcommand{\AppSqueez}{Appendix~G\xspace}

\begin{document}
\preprint{PHYSICAL REVIEW LETTERS {\bf 108}, 220501 (2012)}
\title{Toward Global Quantum Communication:
Beam Wandering Preserves Nonclassicality}%
\author{D. Yu. Vasylyev}
\email[E-mail address: ]{DVasylyev@bitp.kiev.ua}
\affiliation{Institut f\"ur Physik, Universit\"at Rostock,
Universit\"atsplatz 3, D-18051 Rostock, Germany}
 \affiliation{Bogolyubov Institute for Theoretical Physics, NAS of Ukraine, Vul.
Metrologichna 14-b, 03680 Kiev, Ukraine}

\author{A. A. Semenov}
\email[E-mail address: ]{sem@iop.kiev.ua}
\affiliation{Institut f\"ur Physik, Universit\"at Rostock,
Universit\"atsplatz 3, D-18051 Rostock, Germany}%
\affiliation{Bogolyubov Institute for Theoretical Physics, NAS of
Ukraine, Vul. Metrologichna 14-b, 03680 Kiev, Ukraine}
\affiliation{Institute of  Physics, NAS of Ukraine, Prospect Nauky
46, 03028 Kiev, Ukraine}

\author{W. Vogel}%
\email[E-mail address: ]{werner.vogel@uni-rostock.de}
\affiliation{Institut f\"ur Physik, Universit\"at Rostock,
Universit\"atsplatz 3, D-18051 Rostock, Germany}%

\date{\today}

\begin{abstract}
Tap-proof long-distance quantum communication requires a deep
understanding of the strong losses in transmission channels. Here we
provide a rigorous treatment of the effects of beam wandering, one
of the leading disturbances in atmospheric channels, on the quantum
properties of light. From first principles we derive the probability
distribution of the beam transmissivity, with the aim to completely
characterize the quantum state of light. It turns out that beam
wandering may preserve nonclassical effects, such as entanglement,
quadrature and photon number squeezing, much better than a standard
attenuating channel of the same losses.
\end{abstract}

\maketitle

\paragraph*{Introduction.--}

The quantum properties of light open challenging perspectives for
secure data transfer~\cite{Gisin}. Quantum communication is already
used commercially, but its application is still restricted by
relatively small distances. The main limitations are the strong
losses over long distances. For example, in optical fibers typical
values are $0.2\,{\rm dB}/{\rm km}$~\cite{Mitschke}, allowing us to
preserve quantum entanglement over $100\,{\rm km}$~\cite{Huebel}.
However, even this distance is insufficient for global quantum
communication. Some possibilities to improve this situation are
based on protocols, which are rather stable against
losses~\cite{Takesue}, or on the idea of quantum
repeaters~\cite{Briegel}.

A promising alternative is based on free-space channels for
satellite-based global quantum communication, which can bridge any
distance on the Earth. In recent experiments quantum entanglement of
light has been demonstrated, after transmission through a free-space
channel of $144\,{\rm km}$ with atmospheric losses of $32\,{\rm
dB}$~\cite{Ursin, Fedrizzi}. Atmospheric losses are instable and
permanently fluctuating. The related noise is expected to
substantially diminish the quantum properties of the transmitted
light. However, such fluctuations may even be useful to preserve
quantum entanglement~\cite{Semenov2010}.

Noise effects in free-space channels have been extensively
studied~\cite{Tatarskii, Fante}, including models with random
fluctuations and stochastic modulation of the
intensity~\cite{Jakeman, Andrews, Diament, Esposito}. More recently,
the method of fluctuating-loss channels has been
introduced~\cite{Semenov2009}, based on fluctuations of the complex
transmission coefficient, whose absolute value is bounded by
one~\cite{Dong, Shapiro}. Beyond the earlier models~\cite{Jakeman,
Andrews, Diament}, the latter work in Ref.~\cite{Semenov2009} yields
a consistent description of quantum channels.

The probability distribution of the transmission coefficient (PDTC)
is the main characteristic of fluctuating-loss channels. However,
the lack of a consistent model for the PDTC prevents us from
exploiting the advantages of this theory under realistic conditions.
The phenomenological log-normal model was studied for the random
intensity modulation~\cite{Diament}, but it incorrectly describes
the distribution tail with the transmission coefficient being close
to 1. This domain, however, is most important for preserving the
quantum effects of light needed for tap-proof communication.

For a realistic description of the quantum effects of light in
free-space channels, the derivation of a PDTC from first principles
is indispensable. Among the unwanted disturbances in free-space
channels we will concentrate on the important phenomenon of beam
wandering, caused by turbulence~\cite{Tatarskii, Fante} and unstable
adjustment of the radiation source. We obtain an analytical
expression for the transmission coefficient of an aperture. Based on
a normal distribution of the beam positioning~\cite{Esposito}, we
derive an analytical form for the corresponding PDTC. Later on, we
study the effects of beam wandering on quantum phenomena, such as
photon-number and quadrature squeezing, and violations of Bell
inequalities. We show that despite of the extra noise, fluctuating
loss channels may preserve such nonclassical properties much better
than similar constant-loss channels. This result gives a strong
argument to favor free-space quantum communication.

\paragraph*{Fluctuating-loss channels.--} The description of
losses in linear quantum optics usually connects annihilation
operators of the input and output fields, $\hat a_{\rm in}$ and
$\hat a_{\rm out}$, by the standard input-output relation, $\hat
a_{\rm out}=T\hat a_{\rm in}+\sqrt{1-T^2}\hat c$
cf.~\cite{Supplementary}, \AppOIOR. The operator $\hat c$ describes
modes of the environment being in the vacuum state. For our further
considerations we assume the absence of dephasing. This applies to
many experiments, such as photodetection~\cite{Diament} and
polarization analysis for testing Bell inequalities~\cite{Fedrizzi}.
One may also perform homodyning, using a local oscillator
copropagating with the signal~\cite{Elser, Semenov2011}. Hence, the
transmission coefficient $T$ is real and positive. Additionally,
preserving the commutation relations implies that
$T{\in}\left[0,1\right]$. In fluctuating-loss channels, $T$ is a
random variable. The corresponding quantum-state input-output
relation attains its simplest form in the Glauber-Sudarshan $P$
representation~\cite{Semenov2009} (cf.~\cite{Supplementary},
\AppQSIOR),
\begin{equation}
P_{\rm out}(\alpha)=\int_0^1 \ud T\, \mathcal{P}(T)\frac{1}{T^2}
P_{\rm in}\left(\frac{\alpha}{T}\right),\label{IOR}
\end{equation}
where $P_{\rm in}\left(\alpha\right)$ and $P_{\rm
out}\left(\alpha\right)$ are the input and output $P$~functions,
respectively. Fluctuations of losses are described by the PDTC,
$\mathcal{P}(T)$, which we will derive for the scenario of beam wandering.

\begin{figure}[h!]
\includegraphics{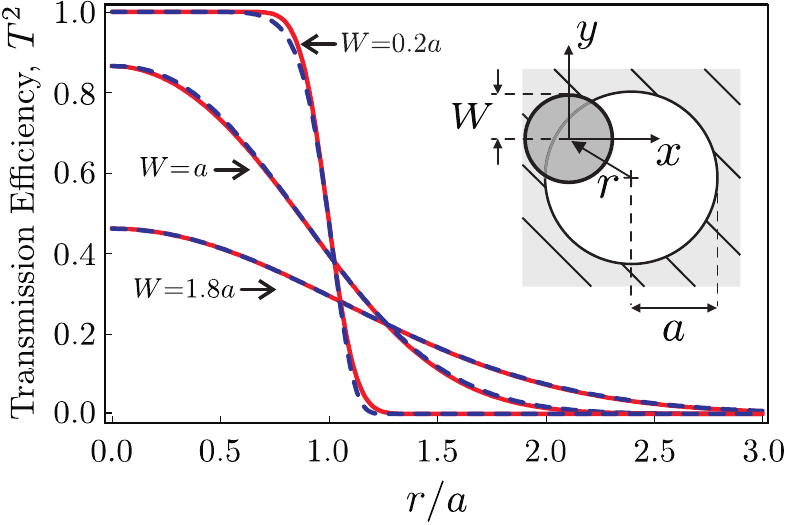}
\caption{\label{fig:T2}  (color online) Transmission efficiency,
$T^2$, vs the beam-deflection distance, $r$, for different values of
the beam-spot radius $W$. Solid and dashed lines correspond to the
numerically calculated integral~(\ref{QWeber}) and the analytical
approximation~(\ref{Efficiency}), respectively. The inset
schematically shows the aperture of radius $a$ and the beam-spot of
radius $W$.}
\end{figure}

\paragraph*{Aperture transmission coefficient.--} For weak absorption,
beam-wandering losses are dominant. Typically they are caused by
aperture truncation of the light at the receiver. This implies that
$T$ in Eq.~(\ref{IOR}) is the aperture transmission coefficient of a
light pulse. We consider the pulse as a superposition of Gaussian
beams~\cite{Mandel} with different wave numbers $k$, propagating
along the $Z$~axis onto the aperture plane at distance
$z_\mathrm{ap}$ from the source. If beam deflection is caused by
imperfect adjustment of the radiation source or by weak turbulence,
the beam incidence is normal to the aperture plane to a good
approximation. The beam center is deflected by the distance $r$ from
the aperture center, see the inset in Fig.~\ref{fig:T2}.

The transmission efficiency (square of the transmission coefficient) of
a Gaussian beam reads as
\begin{equation}
T^2(k)=\int_{\mathcal A}\ud x\, \ud
y\left|U(x,y,z_\mathrm{ap};k)\right|^2,\label{TUin1}
\end{equation}
where $\mathcal A$ is the area of aperture opening,
$U(x,y,z_\mathrm{ap};k)$ is the Gaussian beam normalized in the $XY$
plane. In practice the transmission coefficient $T(k)$ approximately
equals the pulse transmission coefficient in Eq.~(\ref{IOR}). That
is, the pulse spectrum is irrelevant, $T\approx T(k_0)$, with $k_0$
being the carrier wave-number. For a Gaussian beam with the spot
radius $W$ at the aperture plane the transmission efficiency is
given by the incomplete Weber integral~\cite{Agrest},
\begin{equation}
T^2=\frac{2}{\pi W^2}e^{-2\frac{r^2}{W^2}} \int_{0}^a
\ud\varrho\,\varrho\,
e^{-2\frac{\varrho^2}{W^2}}\,\I_0\Bigl(\frac{4}{W^2}r\varrho\Bigr),
\label{QWeber}
\end{equation}
$a$ is the aperture radius and $\I_n$ the modified Bessel function.
Details can be found in \cite{Supplementary}, \AppGB.

The integral~(\ref{QWeber}) can be calculated
numerically~\cite{Esposito}. However, for the evaluation of the PDTC
we propose an approximate analytical expression of the form
(cf.~\cite{Supplementary}, \AppWeber)
\begin{equation}
 T^2=T_0^2 \exp\Bigl[-\Bigl(\frac{ r}{R}\Bigr)^\lambda\Bigr],
 \label{Efficiency}
\end{equation}
where $T_0$ is the maximal transmission coefficient for the given
beam-spot radius $W$; $\lambda$ and $R$ are the shape and scale
parameter, respectively. The constants $T_0$, $\lambda$, and $R$ are
obtained from the following procedure. We consider the transmission
efficiency $T^2$ as a function $T^2(r)$ of the beam-deflection
distance $r$. For the integral~(\ref{QWeber}) one can calculate
analytically the particular values $T^2(0)$, $T^2(a)$, and $(\ud
T^2(r)/\ud r)_{r=a}$ cf.~\cite{Agrest}. From the condition
$T^2(0)=T_0^2$ one gets
\begin{equation}
T_0^2 =1-\exp\Bigl[-2\frac{a^2}{W^2}\Bigr].
\end{equation}
The shape and scale parameters, $\lambda$ and $R$, are obtained from
two algebraic equations for the values of $T^2$ and its derivative
at the point $a$,
\begin{equation}
\begin{split}
\lambda =&8\frac{a^2}{W^2} \frac{\exp\bigl[-4\frac{a^2}{W^2}\bigr]
{ \I}_1\bigl(4\frac{a^2}{W^2}\bigr)}{1-\exp[-4 \frac{a^2}{W^2}]{
\I}_0\bigl(4\frac{a^2}{W^2}\bigr)}\\
&\times\Bigl[\ln\Bigl(\frac{2T_0^2}{1-\exp[-4 \frac{a^2}{W^2}]
{ \I}_0\bigl(4\frac{a^2}{W^2}\bigr)}\Bigr)\Bigr]^{-1},
\end{split}
\end{equation}
\begin{equation}
\label{lambda}
 R=a \Bigl[\ln\Bigl(\frac{2T_0^2}{1-\exp[-4 \frac{a^2}{W^2}]
{
\I}_0\bigl(4\frac{a^2}{W^2}\bigr)}\Bigr)\Bigr]^{-\frac{1}{\lambda}}.
\end{equation}
In Fig.~\ref{fig:T2} we compare the numerical result for $T^2(r)$
with the approximation~(\ref{Efficiency})-(\ref{lambda}). The
maximal relative mean quadratic error appears for $W{=}0.23 a$ and
is $1.85\%$. This gives a reasonable accuracy for the applied
technique.

\paragraph*{Probability distribution of the transmission coefficient.--}
We suppose that the beam-center position is normally distributed
with variance $\sigma^2$ around a point at the distance $d$  from
the aperture center. In the case of beam wandering caused by
imperfect adjustment of the radiation source this variance is
$\sigma^2\approx\sigma^2_\vartheta z^2$, where $\sigma^2_\vartheta$
is the variance of the source-deflection angle. Likewise, in the
case of weak atmospheric turbulence, $\sigma^2\approx1.919\, C_n^2
z^3 (2 W_0)^{-1/3}$~\cite{Fante}, where $C_n^2$ is the
index-of-refraction structure constant, $W_0$ is the beam-spot
radius at the radiation source; for more recent results
see~\cite{Berman}. Generally, both variances should be added up.

Based on the above assumption, the beam-deflection distance $r$ fluctuates
according to the Rice distribution~\cite{Jakeman} with the parameters $d$ and
$\sigma$. Provided that the transmission coefficient $T$ is approximated by
Eq.~(\ref{Efficiency}), the PDTC is given by the log-negative generalized Rice
distribution,
\begin{equation}
\begin{split}
\mathcal{P}(T)&{=}\frac{2 R^2}{\sigma^2 \lambda T}\Bigl(2\ln
\frac{T_0}{T}\Bigr)^{\frac{2}{\lambda}{-}1} \I_0\Bigl(\frac{R d}
{\sigma^2}\Bigl[2\ln \frac{T_0}{T}\Bigr]^\frac{1}{\lambda}\Bigr)\\
&\qquad\times\exp\Bigl[-\frac{1}{2\sigma^2}
\Bigl\{R^2\Bigl(2\ln \frac{T_0}{T}\Bigr)^\frac{2}{\lambda}{+}
d^2\Bigr\}\Bigr]\label{RicePDTC},
\end{split}
\end{equation}
for $T\in [0, T_0]$, and $\mathcal{P}(T)=0$ else. In the particular case when the beam fluctuates
around the aperture center, $d{=}0$, this distribution reduces to the log-negative
Weibull distribution,
\begin{equation}
\begin{split}
\mathcal{P}(T){=}&\frac{2 R^2}{\sigma^2 \lambda T}
\Bigl(2\ln
\frac{T_0}{T}\Bigr)^{\frac{2}{\lambda}{-}1}\exp\Bigl[-\frac{1}{2\sigma^2}R^2\Bigl(2\ln
\frac{T_0}{T}\Bigr)^\frac{2}{\lambda}\Bigr],\label{Weibull}
\end{split}
\end{equation}
for $T\in [0, T_0]$, and $\mathcal{P}(T)=0$ else. More details about
properties of the PDTC can be found in \cite{Supplementary},
\AppPDTC.

\begin{figure}[h!]
\includegraphics{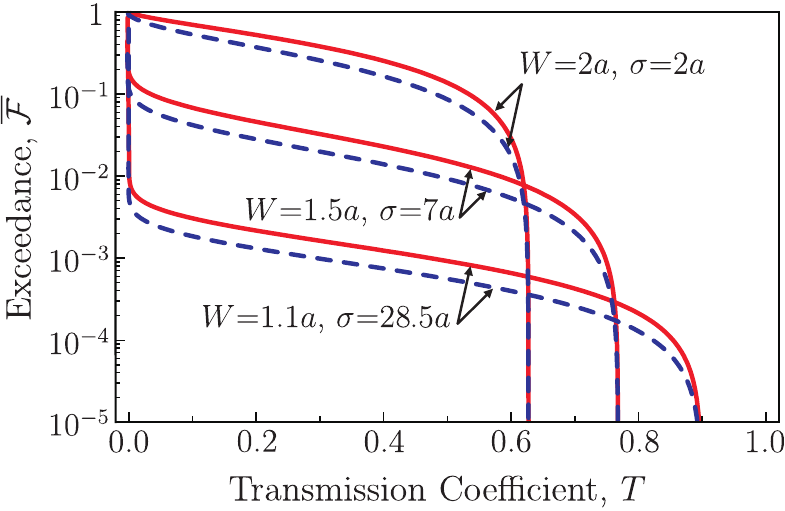}
\caption{\label{fig:Exceedance} (color online) The exceedance (tail
distribution) cf.~Eq.~(\ref{Exceedance}), is shown in the
logarithmic scale for different values of the beam-spot radius $W$,
and standard deviation of beam deflection $\sigma$. Solid and dashed
lines correspond to $d{=}0$ and $d{=}\sigma$, respectively.}
\end{figure}

The PDTC~(\ref{RicePDTC}) and (\ref{Weibull}) may have singularities
at the points $T{=}0$ and $T{=}T_0$. Evidently, in any experimental
reconstructions of the PDTC this singularity cannot be observed
directly due to finite values of the sampling-data number and the
sampling interval for $T$. For this reason, we will use the
exceedance (tail distribution),
\begin{equation}
{\mathcal{\overline{F}}}\!\left(T\right)=\int_{T}^1\ud T^\prime\,
\mathcal{P}\!\left(T^\prime\right),\label{Exceedance}
\end{equation}
which is the probability that the transmission coefficient exceeds
the value $T$. Plots of this function are presented in
Fig.~\ref{fig:Exceedance}. In the following we will show that
contributions from tails with large $T$ are important for preserving
nonclassical properties of the transmitted light. This implies that
the exceedance may also characterize the feasibility of quantum
protocols.

\paragraph*{Nonclassical effects.--}
Input-output relation~(\ref{IOR}) and PDTC~(\ref{RicePDTC}),
(\ref{Weibull}) enable to consistently characterize quantum states
of the transmitted light. We will use this possibility for studying
disturbances of nonclassical effects by beam wandering. In this
context the fluctuating-loss channel with the mean transmission
efficiency $\left\langle T^2\right\rangle$ is compared with a
standard attenuating channel with the transmission efficiency
$T^2{=}\left\langle T^2\right\rangle$.

\begin{figure}[h!]
\includegraphics{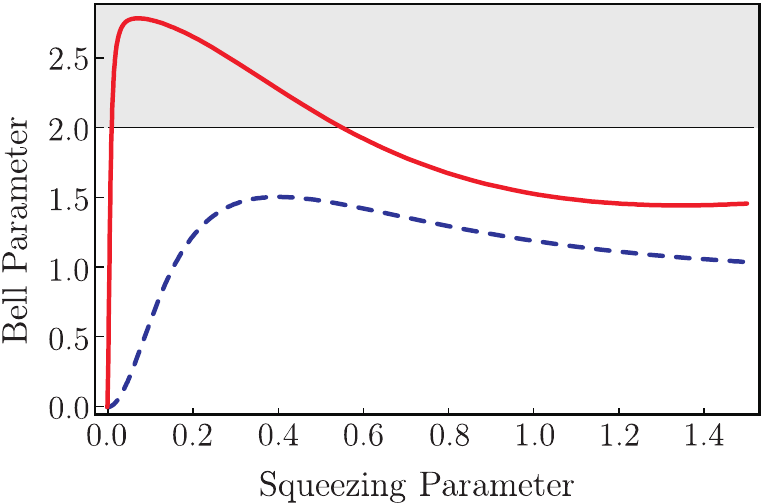}
\caption{\label{fig:BellB}  (color online) The Bell parameter is
shown in dependence of the squeezing  parameter of a parametric
down-conversion source. The solid line shows the situation for beam
wandering, with a standard deviation $\sigma=28.5 a$ of beam
deflection and a beam-spot radius of $W=1.1 a$, leading to a mean
loss of $32\,{\rm dB}$. The receiver-module losses are supposed to
be $9\,{\rm dB}$. The dashed line shows the behavior for constant
losses of the same size. The mean number of stray-light and dark
counts is assumed to be $10^{-5}$. The shaded area indicates the
violation of the Bell inequality.}
\end{figure}

A remarkable example is the experimental violation of Bell
inequalities, after transmission over a $144\,{\rm km}$ atmospheric
channel with a mean loss of $32\,{\rm dB}$~\cite{Fedrizzi}. As it
has been reported in Ref.~\cite{Semenov2010}, this result is very
sensitive to stray-light and dark counts. For example, such a
violation is impossible for constant attenuation of the same size
and the mean number of noise counts of $10^{-5}$ (instead of
$0.5{\times}10^{-6}$ in Ref.~\cite{Fedrizzi}) cf. the dashed line in
Fig.~\ref{fig:BellB}. Indeed, in this case the required simultaneous
clicks of the detectors are mainly caused by stray-light and dark
counts. For fluctuating losses events with large $T$, i.e. in the
tail of the PDTC, ensure the simultaneous clicks of detectors from
an entangled-photon source. A problem is that the earlier models,
e.g. the log-normal PDTC, do not give a correct behavior for the
tail.

Hence it is important to verify that the tail of a consistent PDTC
really improves violations of Bell inequalities, compared with the
case of constant losses. In the present consideration we suppose
that (on average) $32\,{\rm dB}$ of fluctuating losses are caused by
beam wandering only. In fact, other phenomena significantly
contribute to the measured statistics as well~\cite{Ursin}. Here we
effectively model all of them by the PDTC~(\ref{Weibull}). We also
include into consideration $9\,{\rm dB}$ losses of the receiver
module.

We provide calculations similar to Ref.~\cite{Semenov2010} but for
the PDTC~(\ref{Weibull}) cf. also \cite{Supplementary}, \AppBell.
The result is given in Fig.~\ref{fig:BellB}. It clearly shows that
the entanglement can survive in case of beam wandering. Under
similar conditions it disappears for constant losses of the same
size. This finding is of vital importance for long-distance quantum
communication based on entanglement and discrete-variable coding.

The need to observe clicks at two polarization analyzers
post-selects also events with large $T$ in the Bell-inequality test.
In continuous-variable quantum communication based on squeezed
states~\cite{Scarani}, such a natural procedure does not occur. In
order to preserve quadrature squeezing we should additionally
monitor the transmission coefficient and post-select the events with
$T$ exceeding a certain value $T_\mathrm{min}$. Likewise, similar
procedure can be applied for improving the photon-number squeezing.

Monitoring the transmission coefficient can be performed with the
technique proposed in Ref.~\cite{Semenov2011}, where a test pulse
(also used as a local oscillator~\cite{Elser}) is sent in the mode
orthogonally-polarized to the signal. By appropriately choosing
$T_\mathrm{min}$ we can, in principle, preserve any value of
squeezing. However, this is limited by the occurrence of the
corresponding events. Hence the feasibility of this procedure must
be analyzed.

\begin{figure}[h!]
\includegraphics{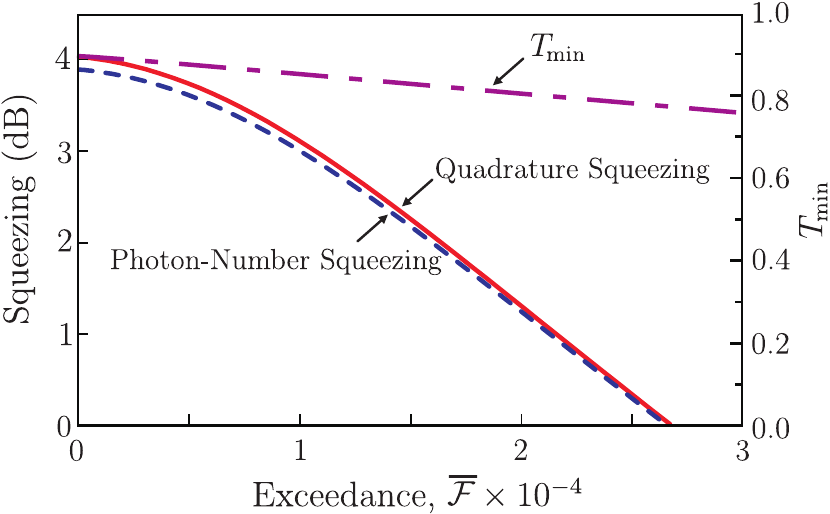}
\caption{\label{fig:Squeezing} (color online) The reachable values
of quadrature and photon-number squeezing are shown as a function of
the corresponding exceedance. The input light is amplitude-squeezed
by $6\,\mathrm{dB}$ and coherently displaced by an amplitude of
$10$. The mean losses are $32\,\mathrm{dB}$ with the same parameters
of beam wandering as in Fig.~\ref{fig:BellB}. The dependence of
$T_\mathrm{min}$ on the exceedance is also shown.}
\end{figure}

To be definite, assume that the input light is amplitude-squeezed by
$6\,\mathrm{dB}$ and coherently displaced by an amplitude of $10$.
Similar to Bell inequality test, we suppose that beam wandering
leads to $32\,\mathrm{dB}$ mean losses. In standard attenuation
channels such losses result in negligibly small squeezing:
$8.8{\times}10^{-4}\,\mathrm{dB}$ for the quadrature and
$8.6{\times}10^{-4}\,\mathrm{dB}$ for the photon number. In
Fig.~\ref{fig:Squeezing} we present the value of squeezing in
dependence on the exceedance~(\ref{Exceedance}) for the point
$T_\mathrm{min}$, for details of calculations
cf.~\cite{Supplementary}, \AppSqueez. It is clearly seen from the
plots that approximately one event from $10^4$ is characterized by
the high level of squeezing. Hence, under such conditions, the
technique is still feasible. This conclusion is of importance for
long-distance quantum communication based on continuous-variable
coding.

\paragraph*{Summary and Conclusions.--} Starting from the input-output relations
for general quantum states we consistently describe the quantum
properties of light undergoing beam wandering in free-space
channels. From first principles we have analytically derived the
main characteristics of such channels -- the probability
distribution of the aperture transmission coefficient. It is given
by the log-negative generalized Rice distribution. When the centers
of beam-wandering and aperture coincide, this function reduces to
the log-negative Weibull distribution. The proposed distribution
does not include phenomena other than beam wandering, which also may
occur in turbulent media. However, our approach is the first model
of free-space channels which consistently applies to the quantum
domain. We believe that further generalizations will be possible.

It has been shown that fluctuating losses associated with beam
wandering can be more favorable for transferring quantum properties
of light, compared with constant losses of the same size, e.g. in
fibers. This includes phenomena such as entanglement, photon-number
and quadrature squeezing. Their conservation plays a crucial role
for free-space quantum communication based on discrete or continuous
variables.

The authors are grateful to Jeffrey H. Shapiro, Alessandro Fedrizzi,
and Jan Sperling for useful comments. This work was partly supported
by the Deutsche Forschungsgemeinschaft through SFB 652.

\section*{Supplemental Material}

\makeatletter \@addtoreset{figure}{section} \makeatother
 \renewcommand{\thefigure}{%
 \thesection\arabic{figure}}

\appendix

\section{Operator input-output relation}
\label{App:OpIOR}

The description of linear losses in the absence of dephasing is
usually based on the operator input-output relation,
\begin{equation}
 \hat{a}_\mathrm{out}=T\hat{a}_\mathrm{in}+\sqrt{1-T^2}\hat{c},
\label{IOR_op}
\end{equation}
where $\hat{a}_\mathrm{in}$ and $\hat{a}_\mathrm{out}$ are the input
and output field-annihilation operators, respectively, $\hat{c}$ is
the operator of environment modes, $T\in\left[0,1\right]$ is the
transmission coefficient. Moreover, $T$ is assumed to be a random
variable in the case of fluctuating-loss channels.
Expression~(\ref{IOR_op}) can also be considered as the input-output
relation in the Heisenberg picture of motion. Here we will show how
to derive this relation from the first principles for free-space
channels and to get an explicit form for the transmission
coefficient.

We suppose that the atmosphere absorption is negligible in the
channel. In this case the light is scattered, so that parts of it do
not reach the receiver aperture, which leads to losses. The
positive-frequency part of the field operators before the aperture
(the input field), $\hat{A}_\mathrm{in}\!\left({\bf r},t\right)$,
and after the aperture (the output field),
$\hat{A}_\mathrm{out}\!\left({\bf r},t\right)$, satisfies the wave
equation,
\begin{equation}
 n^2\!\left({\bf r}\right)\frac{\partial^2}{\partial t^2}
\hat{A}_\mathrm{in/out}\!\left({\bf r},t\right)- c^2\Delta
\hat{A}_\mathrm{in/out}\!\left({\bf r},t\right)=0, \label{WaveEq}
\end{equation}
where $c$ is the speed of light in the vacuum and
\begin{equation}
n\!\left({\bf r}\right)=1+\delta n\!\left({\bf
r}\right)\label{IndexOfRefraction}
\end{equation}
is the atmosphere index of refraction. These fields are related to
each other as
\begin{equation}
\hat{A}_\mathrm{out}\!\left({\bf
r},t\right)=\int\limits_{-\infty}^{+\infty}\D^3{\bf
r^\prime}\,T\!\left({\bf r^\prime}, {\bf r}
\right)\,\hat{A}_\mathrm{in}\!\left({\bf
r^\prime},t\right)+\hat{C}\!\left({\bf r},t\right),\label{IOR_pos}
\end{equation}
where $T\!\left({\bf r^\prime}, {\bf r} \right)$ is the aperture
transmission function, $\hat{C}\!\left({\bf r},t\right)$ is the
noise operator associated with the field scattered by the aperture.

Let the input field be initially prepared in the form of a pulse,
which corresponds to the normalized nonmonochromatic mode
${U}_\mathrm{in}\!\left({\bf r},t\right)$. After passing the
aperture this pulse transforms to the form
\begin{equation}
\int\limits_{-\infty}^{+\infty}\D^3{\bf r^\prime}\,T\!\left({\bf
r^\prime}, {\bf r} \right)\,{U}_\mathrm{in}\!\left({\bf
r^\prime},t\right).\label{OutPulseNonN}
\end{equation}
The normalization of this pulse gives the corresponding output
nonmonochromatic mode,
\begin{equation}
{U}_\mathrm{out}\!\left({\bf r},t\right)=\frac{
{\displaystyle\int\limits_{-\infty}^{+\infty}}\D^3{\bf
r^\prime}\,T\!\left({\bf r^\prime}, {\bf r}
\right)\,{U}_\mathrm{in}\!\left({\bf r^\prime},t\right)}
{\left({\displaystyle\int\limits_{-\infty}^{+\infty}}\D^3{\bf
r}\left|{\displaystyle\int\limits_{-\infty}^{+\infty}}\D^3{\bf
r^\prime}\,T\!\left({\bf r^\prime}, {\bf r}
\right)\,{U}_\mathrm{in}\!\left({\bf r^\prime},t\right)\right|^2
\right)^\frac{1}{2}},\label{OutPulseN}
\end{equation}
which is detected and analyzed at the receiver. The functions
${U}_\mathrm{in/out}\!\left({\bf r},t\right)$ also satisfy the wave
equation~(\ref{WaveEq}).

The operators $\hat{A}_\mathrm{in/out}\!\left({\bf r},t\right)$ can
be expanded into series of the nonmonochromatic modes,
\begin{eqnarray}
\hat{A}_\mathrm{in/out}\!\left({\bf r},t\right)=
\hat{a}_\mathrm{in/out}&&{U}_\mathrm{in/out}\!\left({\bf
r},t\right)\label{expan}\\&&+\sum\limits_{n=1}^{+\infty}
\hat{a}_\mathrm{in/out}^{(n)}{U}_\mathrm{in/out}^{(n)}\!\left({\bf
r},t\right),\nonumber
\end{eqnarray}
where ${U}_\mathrm{in/out}^{(n)}\!\left({\bf r},t\right)$ form a
basis of the orthogonal complement of a mode
${U}_\mathrm{in/out}\!\left({\bf r},t\right)$, and
$\hat{a}_\mathrm{in/out}^{(n)}$ are corresponding field-annihilation
operators. Let us substitute Eq.~(\ref{expan}) in
Eq.~(\ref{IOR_pos}). Utilizing the orthogonality condition,
\begin{equation}
\int\limits_{-\infty}^{+\infty}\D^3{\bf
r}\,{U}_\mathrm{out}^{{(n)}\ast}\!\left({\bf
r},t\right){U}_\mathrm{out}\!\left({\bf r},t\right)=0,
\end{equation}
mode normalization, and Eq.~(\ref{OutPulseN}), we get operator
input-output relation~(\ref{IOR_op}) with the transmission
coefficient,
\begin{equation}
T=\left(\int\limits_{-\infty}^{+\infty}\D^3{\bf
r}\left|\int\limits_{-\infty}^{+\infty}\D^3{\bf
r^\prime}\,T\!\left({\bf r^\prime}, {\bf r}
\right)\,{U}_\mathrm{in}\!\left({\bf r^\prime},t\right)\right|^2
\right)^\frac{1}{2},\label{T_pulse}
\end{equation}
being the normalization constant for the output pulse, cf.
Eqs.~(\ref{OutPulseNonN}) and (\ref{OutPulseN}). All terms
containing the field-annihilation operators
$\hat{a}_\mathrm{in}^{(n)}$ are included in the noise operator
$\hat{c}$ of the input-output relation~(\ref{IOR_op}). The
corresponding modes also contribute to the output pulse, however,
all of them are in the vacuum state.

\section{Quantum-state input-output relations}
\label{App:QSIOR}

In this Appendix we discuss the transformation of the operator
input-output relation~(\ref{IOR_op}) to the Schr\"odinger picture of
motion, to obtain the corresponding density ope\-rators. The easiest
way to solve this problem is based on the Glauber-Sudarshan $P$
representation. In this representation the quantum-state
input-output relation resembles the corresponding relation in
classical optics.

Let us remind that the $P$ function, $P\!\left(\alpha\right)$, is
related to the corresponding density operator $\hat{\varrho}$ as
\begin{equation}
P\!\left(\alpha\right)=\frac{1}{\pi^2}
\int\limits_{-\infty}^{+\infty}\D^2\beta\,
\Phi\!\left(\beta\right)e^{\alpha\beta^*-\alpha^*\beta},
\label{PFunc}
\end{equation}
where
\begin{equation}
 \Phi\!\left(\beta\right)=\Tr\left[\hat{\varrho}\,
\exp\left(\hat{a}^{\dag}\beta\right)\exp\left(-\hat{a}\beta^\ast\right)
\right] \label{CharQuantP}
\end{equation}
is the corresponding characteristic function. By using
Eq.~(\ref{IOR_op}), this allows one to derive the relation between
the characteristic function of the attenuated and the input field,
$\Phi_T\!\left(\beta\right)$ and
$\Phi_\mathrm{in}\!\left(\beta\right)$, respectively, as
\begin{equation}
 \Phi_T\!\left(\beta\right)=\Phi_\mathrm{in}\!\left(T\beta\right).
\label{IOR_ChF_T}
\end{equation}
Here we utilize the fact that the characteristic function of the
environment modes is equal to unity since all these modes are
assumed to be in the vacuum state. Substitution of
Eq.~(\ref{IOR_ChF_T}) in Eq.~(\ref{PFunc}) gives the input-output
relation between the corresponding $P$ functions,
\begin{equation}
 P_T\!\left(\alpha\right)=\frac{1}{T^2}
P_\mathrm{in}\!\left(\frac{\alpha}{T}\right).\label{IOR_P_T}
\end{equation}
Now we have to take into account the fact that the transmission
coefficient $T$ fluctuates randomly. Consequently, we have to
average relations~(\ref{IOR_ChF_T}) and (\ref{IOR_P_T}) with the
probability distribution of the transmission coefficient (PDTC),
$\mathcal{P}\!\left(T\right)$. The resulting input-output relations
read as
\begin{equation}
 \Phi_\mathrm{out}\!\left(\beta\right)=
\int\limits_{0}^{1}\D T \mathcal{P}\!\left(T\right)
\Phi_\mathrm{in}\!\left(T\beta\right),\label{IOR_ChF}
\end{equation}
 \begin{equation}
 P_\mathrm{out}\!\left(\alpha\right)=
\int\limits_{0}^{1}\D T \mathcal{P}\!\left(T\right) \frac{1}{T^2}
P_\mathrm{in}\!\left(\frac{\alpha}{T}\right),\label{IOR_P}
\end{equation}
where $P_\mathrm{out}\!\left(\alpha\right)$ and
$\Phi_\mathrm{out}\!\left(\beta\right)$ are the $P$ function and the
corresponding characteristic function, respectively, of the output
field.

The quantum-state input-output relation~(\ref{IOR_P}) can be easily
rewritten in any other representations. For example, the
corresponding Wigner functions, $W_\mathrm{in}\!\left(\alpha\right)$
and $W_\mathrm{out}\!\left(\alpha\right)$, are related to each other
as
\begin{widetext}
\begin{equation}
W_\mathrm{out}\!\left(\alpha\right)= \int\limits_{0}^{1}\D T
\mathcal{P}\!\left(T\right) \int\limits_{-\infty}^{+\infty}\D^2\beta
\frac{2}{\pi T^2 \left(1-T^2\right)}
\exp\left[-\frac{2\left|\alpha-\beta\right|^2}{1-T^2}\right]
W_\mathrm{in}\!\left(\frac{\beta}{T}\right),
\end{equation}
\end{widetext}
which requires an additional $\beta$ integration. It is also useful
to have the corresponding relation for the normal-ordered moments,
$M_{n,m}=\Tr\left(\hat{\varrho}\hat{a}^{\dag n}\hat{a}^m\right)$,
\begin{equation}\label{IOmoments}
M_{n,m}^\mathrm{out}=\left\langle T^{n+m}\right\rangle
M_{n,m}^\mathrm{in},
\end{equation}
which directly leads to Eqs.~(\ref{AppG_pnsq}) and (\ref{AppG_qsq})
of Appendix~\ref{App:Sq}.

\section{Pulsed Gaussian beam}
\label{GaussianBeam}

In this Appendix we specify the form of the transmission
coefficient~(\ref{T_pulse}) for the case of the input pulse prepared
as a superposition of Gaussian beams, see Ref.~\cite{Mandel}, with
different wave numbers $k$,
\begin{eqnarray}
{U}_\mathrm{in}\!&&\left({\bf
r},t\right)\label{SuperposGaussBeams}\\&&=\frac{1}{\sqrt{2\pi}}
\int\limits_{-\infty}^{+\infty}\D k\,
C\!\left(k\right)S_\mathrm{in}\!\left(x,y,z;
k\right)\exp\!\left[ik\left(z-ct\right)\right],\nonumber
\end{eqnarray}
where $C\!\left(k\right)$ is the pulse spectrum,
$S_\mathrm{in}\!\left(x,y,z;k\right)$ is the input-beam envelope.
The beams propagation direction corresponds to the $Z$-axis. Many
laser-based sources emit light, which can be approximately described
by Eq.~(\ref{SuperposGaussBeams}). The non-normalized pulse after
the aperture, cf. Eq.~(\ref{OutPulseNonN}), has the form similar to
Eq.~(\ref{SuperposGaussBeams}),
\begin{eqnarray}
&&\int\limits_{-\infty}^{+\infty}\D^3{\bf r^\prime}\,T\!\left({\bf
r^\prime}, {\bf r}
\right)\,{U}_\mathrm{in}\!\left({\bf r^\prime},t\right)\label{SuperposGaussBeamsOut}\\
&&=\frac{1}{\sqrt{2\pi}}\int\limits_{-\infty}^{+\infty}\D k\,
C\!\left(k\right)S_\mathrm{out}\!\left(x,y,z;
k\right)\exp\!\left[ik\left(z-ct\right)\right],\nonumber
\end{eqnarray}
where $S_\mathrm{out}\!\left(x,y,z; k\right)$ is the output-beam
envelope.

Since the pulses~(\ref{SuperposGaussBeams}) and
(\ref{SuperposGaussBeamsOut}) obey the wave equation~(\ref{WaveEq}),
the input- and output-beam envelopes in the paraxial approximation
satisfy the equation
\begin{equation}
\left[i\frac{\partial }{\partial z}+\frac{1}{2k}\Delta_\bot+k\delta
n\!\left({\bf r}\right)\right]
S_\mathrm{in/out}\!\left(x,y,z;k\right)=0,\label{Paraxial}
\end{equation}
for details see Ref.~\cite{Mandel}. Here $\Delta_\bot$ is the
transverse part of the Laplace operator. This equation preserves the
norm in the $(X,Y)$ plane, i.e. the relation
$\int\limits_{-\infty}^{+\infty}\D x \, \D
y\left|S_\mathrm{in/out}\!\left(x,y,z; k\right)\right|^2$ does not
depend on $z$ and equals to unity for the input pulse. The statement
that the source irradiates Gaussian beams means that
$S_\mathrm{in}\!\left(x,y,z;k\right)$ is the solution of
Eq.~(\ref{Paraxial}), with the following condition at $z=0$:
\begin{eqnarray}
S_\mathrm{in}&&\left(x,y,0;k\right)\label{GaussBeam0}\\&&=\sqrt{\frac{2}{\pi
W_0^2}}\exp\left[-\left(\frac{1}{2W_0^2}{+}\frac{i k}{2R_0}\right)
\left(x^2{+}y^2\right)\right]e^{i\Psi_0},\nonumber
\end{eqnarray}
where  $W_0$, $R_0$, and $\Psi_0$ are values at the transmitter for
the beam-spot radius, curvature radius of the wavefront, and the
Gouy phase, respectively.

We substitute Eq.~(\ref{SuperposGaussBeamsOut}) in
Eq.~(\ref{T_pulse}) and take into account the fact that
$S_\mathrm{out}\!\left(x,y,z; k\right)$ is slowly varying with
respect to $z$, comparing with $\exp\left[ikz\right]$. Hence, in the
corresponding integration we can approximate
$S_\mathrm{out}\!\left(x,y,z; k\right)$ by
$S_\mathrm{out}\!\left(x,y,z_\mathrm{ap}; k\right)$, where
$z_\mathrm{ap}$ is the position of the aperture plane at the
$Z$-axis. This results in the fact that the transmission efficiency
(square of the transmission coefficient) has the form of
decomposition,
\begin{equation}
T^2=\int\limits_{-\infty}^{+\infty}\D
k\left|C\!\left(k\right)\right|^2T^2\!\left(k\right),\label{T2Sout}
\end{equation}
of the beam transmission efficiencies,
\begin{equation}
T^2\!\left(k\right)=\int\limits_{-\infty}^{+\infty}\D x \, \D
y\left|S_\mathrm{out}\!\left(x,y,z_\mathrm{ap};
k\right)\right|^2,\label{T2Beam}
\end{equation}
with different wave numbers $k$.

At the aperture plane, $z=z_\mathrm{ap}$, the input and output beam
envelops are related to each other as
\begin{equation}
S_\mathrm{out}\!\left(x,y,z_\mathrm{ap};
k\right)=\left\{\begin{array}{ll}
S_\mathrm{in}\!\left(x,y,z_\mathrm{ap}; k\right), &\textrm{for}
\left(x,y\right)\in\mathcal{A}\\
0,&\textrm{for}\left(x,y\right)\notin\mathcal{A}
\end{array}
\right. ,
\end{equation}
where $\mathcal{A}$ is the area of aperture opening. This implies
that Eq.~(\ref{T2Beam}) reduces to
\begin{equation}
T^2\!\left(k\right)=\int_\mathcal{A}\D x \, \D
y\left|S_\mathrm{in}\!\left(x,y,z_\mathrm{ap};
k\right)\right|^2.\label{T2Pulse}
\end{equation}
Here the integration is taken over the area of aperture opening.

In practice, the shape of the pulse spectrum,
$\left|C\!\left(k\right)\right|^2$, is much narrower than the shape
of the beam transmission efficiency, $T^2\!\left(k\right)$. Besides,
as it follows from the normalization condition for the input pulse,
\begin{equation}
\int\limits_{-\infty}^{+\infty}\D
k\left|C\!\left(k\right)\right|^2=1.
\end{equation}
This yields that the pulse transmission efficiency~(\ref{T2Sout}) is
approximately equal to the beam transmission
efficiency~(\ref{T2Pulse}), $T^2{\approx} T^2\!\left(k_0\right)$,
where $k_0$ is the carrier wave-number. Remaining that the
Gaussian-beam amplitude is
\begin{equation}
U\!\left(x,y,z; k\right)=S_\mathrm{in}\!\left(x,y,z; k\right)
\exp\!\left(ikz\right),\label{GaussBeamField}
\end{equation}
the transmission efficiency can be equivalently given by
\begin{equation}
T^2\approx T^2\!\left(k_0\right)=\int_\mathcal{A}\D x \, \D
y\left|U\!\left(x,y,z_\mathrm{ap};
k_0\right)\right|^2,\label{T2PulseGauss}
\end{equation}
cf.~Eq.~(2).

Therefore, in order to find the transmission coefficient one has to
resolve Eq.~(\ref{Paraxial}) with the boundary
condition~(\ref{GaussBeam0}). For $\delta n\!\left({\bf
r}\right){=}0$ the solution is obtained by replacing the parameters
$W_0$, $R_0$, and $\Psi_0$ in the boundary
condition~(\ref{GaussBeam0}) with the corresponding values $W$, $R$,
and $\Psi$, which depend on $z$, see Ref.~\cite{Mandel} for details.
We model the effect of $\delta n\!\left({\bf r}\right)$ in
Eq.~(\ref{Paraxial}) by deflection of the beam from the aperture
center, which works for the week turbulence only. Moreover, the beam
can also be deflected due to the imperfect source adjustment. Since
the aperture form is circular, all directions appear to be
equivalent. Hence, we can consider, without loss of generality, the
beam-deflection by the distance $r$ in the direction of $X$-axis.
The obtained solution,
\begin{eqnarray}
S_\mathrm{in}&&\left(x,y,z;k\right)\label{GaussBeamZDeflected}\\
&&=\sqrt{\frac{2}{\pi W^2}}\exp\left\{-\Big[\frac{1}{2W^2}{+}\frac{i
k}{2R}\Big]
\Big[\left(x-r\right)^2{+}y^2\Big]\right\}e^{i\Psi},\nonumber
\end{eqnarray}
should be substituted in Eqs.~(\ref{GaussBeamField}) and
(\ref{T2PulseGauss}). This results in the transmission coefficient
expressed in terms of the incomplete Weber integral, which is
approximately evaluated in Appendix~\ref{App:Weber},
\begin{equation}
T^2{=}\frac{2}{\pi
W^2}e^{-2\frac{r^2}{W^2}}\int\limits_0^ad\rho\,\rho\,
e^{-2\frac{\rho^2}{W^2}}\,{\rm
I}_0\Bigl(4\frac{r\rho}{W^2}\Bigr),\label{T2IncWeber}
\end{equation}
where $a$ is the aperture radius, and $\I_0$ is the modified Bessel
function, cf.~Eq.~(3).

\section{Incomplete Weber integral}
\label{App:Weber}

In this Appendix we give some technical details related to the
approximate analytical evaluation of the integral in
Eq.~(\ref{T2IncWeber}), see also Eq.~(3). First note that the
incomplete Weber integral is defined as, see Ref.~\cite{Agrest},
 \begin{equation}
\widetilde Q_n(x,z)=(2x)^{-n-1}e^x\int\limits_0^z dt\, t^{n+1}
\exp\left(-\frac{t^2}{4x}\right) \I_n(t),\label{WeberIntDef}
\end{equation}
where $\I_n$ is the modified Bessel function. In terms of this
special function Eq.~(\ref{T2IncWeber}) can also be given by
\begin{equation}
T^2=e^{-4\frac{r^2}{W^2}}\widetilde
Q_0\left(2\frac{r^2}{W^2},4\frac{ra}{W^2}\right). \label{AppC1}
\end{equation}
An alternative representation for $\widetilde Q_n(x,z)$ consists in
the Lommel series of the modified Bessel functions. However, this
series converges slowly and for practical purposes one usually uses
the numerical integration.

We derive an analytical approximation for Eqs.~(\ref{T2IncWeber})
and (\ref{AppC1}). For this purpose we consider the transmission
efficiency, $T^2$, as a function $T^2\!\left(r\right)$ of the
variable $r$. This function can be effectively approximated as
\begin{equation}
T^2\left(r\right)=T_0^2\exp\left[-\left(\frac{r}{R}\right)^\lambda
\right],\label{AppC2}
\end{equation}
cf.~Eq.~(4), where the parameters $T_0$, $R$, and $\lambda$ (maximal
transmission coefficient, scale, and shape parameters, respectively)
can be obtained with using some properties of the incomplete Weber
integral.

The maximal transmission coefficient $T_0$ is obtained by comparing
the exact, cf. Eq.~(\ref{T2IncWeber}), and approximate, cf.
Eq.~(\ref{AppC2}), expressions for $T^2\!\left(r\right)$ at the
point $r{=}0$. In this case the integral~(\ref{T2IncWeber}) is
evaluated analytically. Using the straightforward calculations, the
maximal transmission efficiency,
\begin{equation}
T^2_0=1-\exp\left[-2\frac{a^2}{W^2}\right],\label{AppC3}
\end{equation}
cf.~Eq.~(5), is obtained explicitly.

Similar consideration can be applied for scale, $R$, and shape,
$\lambda$, parameters. In this case we compare exact and approximate
expressions for $T^2\!\left(r\right)$ and their derivatives $\D
T^2\!\left(r\right)\!/\D r$ at the point $r{=}a$. The corresponding
integrals for the exact relation~(\ref{AppC1}) are evaluated
explicitly as, cf.~Ref.~\cite{Agrest},
  \begin{equation}
T^2\!\left(a\right)=\frac{1}{2}\Bigl\{1-e^{-4\frac{a^2}{W^2}}\,{\rm
I}_0\Bigl(4\frac{a^2}{W^2}\Bigr)\Bigr\},\label{AppC4}
\end{equation}
\begin{equation}
\begin{split}
\left(\frac{\D T^2(r)}{\D
r}\right)_{r=a}&=\frac{4a}{\pi W^2}e^{-4\frac{a^2}{W^2}}\Bigl\{\widetilde Q_1\Bigl(\frac{2a^2}{W^2},\frac{4a^2}{W^2}\Bigr)\\
&\qquad\qquad\qquad\qquad-\widetilde Q_0\Bigl(\frac{2a^2}{W^2},\frac{4a^2}{W^2}\Bigr)\Bigr\}\\
&=-\frac{4a}{w^2}\,e^{-4\frac{a^2}{W^2}}\,\,{\rm
I}_1\Bigr(4\frac{ra}{W^2}\Bigr).
\end{split}\label{AppC5}
\end{equation}
These expressions can be equated with the same quantities evaluated
from the approximate form~(\ref{AppC2}),
\begin{equation}
T^2\!\left(a\right)=T_0^2\exp\left[-\left(\frac{a}{R}\right)^\lambda
\right],\label{T2(a)}
\end{equation}
\begin{equation}
\left(\frac{\D T^2(r)}{\D r}\right)_{r=a}=
-T_0^2\frac{\lambda}{R}\left(\frac{a}{R}\right)^{\lambda-1}
\exp\left[-\left(\frac{a}{R}\right)^\lambda \right].\label{DT2Dr(a)}
\end{equation}
The obtained system of two algebraic equations for $R$ and $\lambda$
yields
\begin{equation}
\begin{split}
\lambda =&8\frac{a^2}{W^2} \frac{\exp\bigl[-4\frac{a^2}{W^2}\bigr] {
\I}_1\bigl(4\frac{a^2}{W^2}\bigr)}{1-\exp[-4 \frac{a^2}{W^2}]{
\I}_0\bigl(4\frac{a^2}{W^2}\bigr)}\\
&\times\Bigl[\ln\Bigl(\frac{2T_0^2}{1-\exp[-4 \frac{a^2}{W^2}] {
\I}_0\bigl(4\frac{a^2}{W^2}\bigr)}\Bigr)\Bigr]^{-1},
\end{split}
\end{equation}
\begin{equation}
\label{BigR}
 R=a \Bigl[\ln\Bigl(\frac{2T_0^2}{1-\exp[-4 \frac{a^2}{W^2}]
{
\I}_0\bigl(4\frac{a^2}{W^2}\bigr)}\Bigr)\Bigr]^{-\frac{1}{\lambda}},
\end{equation}
cf.~Eqs.~(7) and (8).

\section{Beam-wandering PDTC}

In this Appendix we present some technical details related to
deriving the beam-wandering PDTC, cf. Eqs.~(8) and (9), and discuss
mathematical techniques for operating with this object. The basic
assumption is that the beam-center position, $\left(x_o,y_o\right)$,
randomly fluctuates according to the two-dimensional Gaussian
distribution,
\begin{equation}
p(x_o,y_o;\mathbf{d},\sigma)=\frac{1}{2\pi\sigma^2}
\exp\left[-\frac{(x_o-d_x)^2+(y_o-d_y)^2}{2\sigma^2}\right],
\label{Gaussian}
\end{equation}
where $\sigma$ is the standard deviation of beam deflection and
$\mathbf{d}{=}(d_x,d_y)$ is the mean position of beam center. This
implies that the beam-deflection distance $r{=}\sqrt{x_o^2+y_o^2}$
fluctuates according to the Rice distribution,
\begin{equation}
p(r; d,\sigma)=\frac{r}{\sigma^2}\I_0\left(\frac{r
d}{\sigma^2}\right)\exp\Bigl[-\frac{r^2+d^2}{2\sigma^2}\Bigr],
\label{Rice}
\end{equation}
where $d{=}\sqrt{d_x^2+d_y^2}$ is the distance between aperture and
fluctuation centers, $\I_\mathrm{0}$ is the modified Bessel
function.

The random variables $r$ and $T$ are related to each other via
Eq.~(\ref{AppC2}), cf. also Eq.~(4). This yields that the
probability distribution of the transmission coefficient (PDTC)
reads as
\begin{equation}
\begin{split}
\mathcal{P}(T)&{=}\frac{2 R^2}{\sigma^2 \lambda T}\Bigl(2\ln
\frac{T_0}{T}\Bigr)^{\frac{2}{\lambda}{-}1} {\rm I}_0\Bigl(\frac{R
d}
{\sigma^2}\Bigl[2\ln \frac{T_0}{T}\Bigr]^\frac{1}{\lambda}\Bigr)\\
&\qquad\times\exp\Bigl[-\frac{1}{2\sigma^2} \Bigl\{R^2\Bigl(2\ln
\frac{T_0}{T}\Bigr)^\frac{2}{\lambda}{+}
d^2\Bigr\}\Bigr],
\end{split}
\end{equation}
for $T\in [0, T_0]$, and $\mathcal{P}(T)=0$ else, cf. Eq.~(8). For
this distribution we refer to as the log-negative generalized Rice
distribution. This is also important to note that log-losses
$\theta{=}-\ln T^2$ are distributed according to the generalized
Rice distribution,
\begin{equation}
\begin{split}
\mathcal{P}(\theta)=&\frac{R^2}{\sigma^2\lambda}\bigl(\theta{-}\theta_0\bigr)^{\frac{2}{\lambda}-1}{\rm I}_0\left(\frac{Rd}{\sigma^2}\bigl(\theta{-}\theta_0\bigr)^{\frac{1}{\lambda}}\right)\\
&\times\exp\Bigl[-\frac{1}{2\sigma^2}\Bigl(R^2\bigl(\theta{-}\theta_0\bigr)^{\frac{2}{\lambda}}+d^2\Bigr)\Bigr],
\end{split}\label{DistrTheta}
\end{equation}
for $\theta\ge\theta_0$ and $\mathcal{P}(\theta){=}0$ else, where
$\theta_0{=}{-}\ln T_0^2$. A particular case of
Eqs.~(\ref{RicePDTC}) and (\ref{DistrTheta}) for $d{=}0$ is the
log-negative Weibull distribution for the transmission coefficient,
\begin{equation}
\begin{split}
\mathcal{P}(T){=}&\frac{2 R^2}{\sigma^2 \lambda T} \Bigl(2\ln
\frac{T_0}{T}\Bigr)^{\frac{2}{\lambda}{-}1}\exp\Bigl[-\frac{1}{2\sigma^2}R^2\Bigl(2\ln
\frac{T_0}{T}\Bigr)^\frac{2}{\lambda}\Bigr]
\end{split}
\end{equation}
for $T\in [0, T_0]$, and $\mathcal{P}(T)=0$ else, cf. Eq.~(9), and
the Weibull distribution with the shape parameter
$\left({\sqrt{2}\sigma}/{R}\right)^\lambda$ and the scale parameter
$2/\lambda$ for log-losses,
\begin{equation}
\begin{split}
\mathcal{P}(\theta)=&\frac{R^2}{\sigma^2\lambda}\bigl(\theta{-}\theta_0\bigr)^{\frac{2}{\lambda}-1}
\exp\Bigl[-\frac{1}{2\sigma^2}R^2
\bigl(\theta{-}\theta_0\bigr)^{\frac{2}{\lambda}}\Bigr]
\end{split}\label{DistrTheta1}
\end{equation}
for $\theta\ge\theta_0$ and $\mathcal{P}(\theta){=}0$ else.

Integrations over the PDTC~(\ref{RicePDTC}) and (\ref{Weibull}) is
convenient to perform with respect to the measure $\D r(T)$. Here
\begin{equation}
r(T)=R\left(2\ln\frac{T_0}{T}\right)^\frac{1}{\lambda}\label{r(T)}
\end{equation}
inverts Eq.~(\ref{AppC2}). In practice this means that the mean
value for the function $f(T)$ of the transmission coefficient $T$
can be obtained as
\begin{equation}
\int\limits_0^1\D T\, \mathcal{P}(T)\,f[T]=\int\limits_0^{+\infty}\D
r\, p(r;d,\sigma)\, f[T(r)],
\end{equation}
where $p(r;d,\sigma)$ is the Rice distribution.

Fluctuations of the transmission coefficient can also be
characterized by the cumulative probability distribution,
\begin{equation}
{\mathcal{F}}\!\left(T\right)=\int\limits_{0}^T\D T^\prime\,
\mathcal{P}\!\left(T^\prime\right).\label{CPDTC}
\end{equation}
For our purposes, however, it is convenient to use the exceedance
(tail distribution), which is defined as
\begin{equation}
{\mathcal{\overline{F}}}\!\left(T\right)=1-
{\mathcal{F}}\!\left(T\right)=\int\limits_{T}^1\D T^\prime\,
\mathcal{P}\!\left(T^\prime\right),\label{ExceedanceCum}
\end{equation}
cf. Eq.~(10). Clearly, the exceedance is equal to the probability
that the transmission coefficient exceeds the value of $T$.

The exceedance of log-negative Rice distribution~(\ref{RicePDTC}) is
expressed in terms of the incomplete Weber
integral~(\ref{WeberIntDef}),
\begin{equation}
{\mathcal{\overline{F}}}\!\left(T\right)=
1{-}\exp\left[-\frac{d^2}{\sigma^2}\right]
\widetilde{Q}_0\!\left(\frac{d^2}{2\sigma^2}, \frac{r(T)
d}{\sigma^2}\right),\label{ExceedanceRiceExact}
\end{equation}
where $r\!\left(T\right)$ is given by Eq.~(\ref{r(T)}). For this
special function we can apply the technique presented in
Appendix~\ref{App:Weber}. This results in an approximate expression,
\begin{equation}
\overline{\mathcal F}\!\left(T\right){=}1{-}{\mathcal
F}_0\!\left(T\right)\exp\left[-\Bigl(\frac{d}{D\!\left(T\right)}\Bigr)^
{\mu\!\left(T\right)}\right],\label{ExceedanceAppr}
\end{equation}
where
\begin{equation}
{\mathcal
F}_0\!\left(T\right)=1{-}\exp\left[-\frac{r^2\!\left(T\right)}{2\sigma^2}\right],
\end{equation}
\begin{equation}
D\!\left(T\right)=r\!\left(T\right)\Biggl\{\ln\left[\frac{2{\mathcal
F}_0\!\left(T\right)}{1-{\I}_0\bigl(\frac{r^2\!\left(T\right)}{\sigma^2}\bigr)
\exp\bigl[-\frac{r^2\!\left(T\right)}{\sigma^2}\bigr]}\right]
\Biggr\}^{-\frac{1}{\mu\!\left(T\right)}},
\end{equation}
\begin{equation}
\begin{split}
\mu\!\left(T\right)=&\frac{2r^2\!\left(T\right)}{\sigma^2}
\frac{{\I}_1\bigl(\frac{r^2\!\left(T\right)}{\sigma^2}\bigr)}
{1-{\I}_0\bigl(\frac{r^2\!\left(T\right)}{\sigma^2}\bigr)
\exp\bigl[-\frac{r^2\!\left(T\right)}{\sigma^2}\bigr]}
e^{-\frac{r^2\!\left(T\right)}{\sigma^2}}\\
&\quad\times\Biggl\{\ln\left[\frac{2{\mathcal
F}_0\!\left(T\right)}{1-{\I}_0\bigl(\frac{r^2\!\left(T\right)}
{\sigma^2}\bigr)\exp\bigl[-\frac{r^2\!\left(T\right)}
{\sigma^2}\bigr]}\right]\Biggr\}^{-1}.
\end{split}
\end{equation}
For $d{=}0$ the approximate expression~(\ref{ExceedanceAppr})
reduces to
\begin{equation}
{\mathcal{\overline{F}}}\!\left(T\right)=
1-{\mathcal{{F}}}_0\!\left(T\right)=
\exp\left[-\frac{r^2\!\left(T\right)}{2\sigma^2}\right],\label{ExceedanceWeibull}
\end{equation}
which also coincides with the exact form of the exceedance for the
log-negative Weibull distribution~(\ref{Weibull}).

\section{Bell-inequality test}
\label{App:Bell}

\begin{figure}[ht!]
\includegraphics[clip=,width=\linewidth]{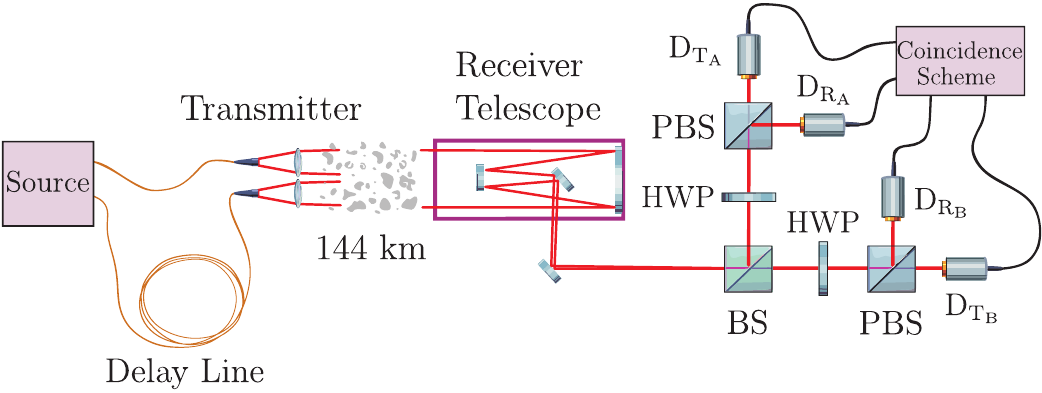}
\caption{\label{Fig:Bell} Setup for checking Bell inequalities with
the light transmitted through a $144\,\mathrm{km}$ atmospheric
channel, cf.~Ref.~\cite{Fedrizzi}. A parametric down-conversion
source produces entangled photon pairs. These photons, separated by
a small time interval, are sent to the receiver. After collecting by
the telescope, photons are split by a beam-splitter $\mathrm{BS}$
and then treated by the corresponding polarization analyzers. Each
polarization analyzer consists of (i) half-wave plate,
$\mathrm{HWP}$, which changes the polarization angles
$\theta_\mathrm{A}$ and $\theta_\mathrm{B}$; (ii) polarizing
beam-splitter, $\mathrm{PBS}$; (iii) on/off detectors
$\mathrm{D}_\mathrm{T_A}$, $\mathrm{D}_\mathrm{T_B}$ for the
transmitted light, and $\mathrm{D}_\mathrm{R_A}$,
$\mathrm{D}_\mathrm{R_B}$ for the reflected light. }
\end{figure}

In this Appendix we remind technical details from
Ref.~\cite{Semenov2010} concerning calculations of the Bell
parameter for the light generated by a parametric down-conversion
source, transmitted through the atmosphere, and then treated by
polarization analyzers with on/off detectors. We consider
experimental configuration used in Ref.~\cite{Fedrizzi}, cf. also
Fig.~\ref{Fig:Bell}. In this scenario both receivers are situated at
the same place and photons are split by a small time interval.

The Bell inequality in the Clauser-Horne-Shimony-Holt (CHSH) form
states that the parameter
\begin{eqnarray}
\mathcal{B}&=&\left|E\!\left(\theta_\mathrm{A}^{(1)},
\theta_\mathrm{B}^{(1)}\right)-E\!\left(\theta_\mathrm{A}^{(1)},
\theta_\mathrm{B}^{(2)}\right)\right|\label{BellParameter}\\
&+&\left|E\!\left(\theta_\mathrm{A}^{(2)},
\theta_\mathrm{B}^{(2)}\right)+E\!\left(\theta_\mathrm{A}^{(2)},
\theta_\mathrm{B}^{(1)}\right)\right|\nonumber,
\end{eqnarray}
also referred to as the Bell parameter, cannot exceed the value of
$2$ in any local realistic theory. Quantum theory violates the Bell
inequality. The correlation coefficients
$E\!\left(\theta_\mathrm{A},\theta_\mathrm{B}\right)$ in
Eq.~(\ref{BellParameter}) are defined as
\begin{equation}
E\left(\theta_\mathrm{A}, \theta_\mathrm{B}\right) =
\frac{P_\mathrm{same}\left(\theta_\mathrm{A},
\theta_\mathrm{B}\right)-P_\mathrm{different}\left(\theta_\mathrm{A},
\theta_\mathrm{B}\right)}{P_\mathrm{same}\left(\theta_\mathrm{A},
\theta_\mathrm{B}\right)+P_\mathrm{different}\left(\theta_\mathrm{A},
\theta_\mathrm{B}\right)},\label{correlation}
\end{equation}
where
\begin{equation}
P_\mathrm{same}\!\left(\theta_\mathrm{A},
\theta_\mathrm{B}\right)=P_\mathrm{\mathrm{T_A},
\mathrm{T_B}}\!\left(\theta_\mathrm{A},
\theta_\mathrm{B}\right)+P_\mathrm{\mathrm{R_A},
\mathrm{R_B}}\!\left(\theta_\mathrm{A},
\theta_\mathrm{B}\right)\label{same}
\end{equation}
is the probability to get clicks on detectors in the same channels
of the polarization analyzers, and
\begin{equation}
P_\mathrm{different}\!\left(\theta_\mathrm{A},
\theta_\mathrm{B}\right)=P_\mathrm{\mathrm{T_A},
\mathrm{R_B}}\!\left(\theta_\mathrm{A},
\theta_\mathrm{B}\right)+P_\mathrm{\mathrm{R_A},
\mathrm{T_B}}\!\left(\theta_\mathrm{A},
\theta_\mathrm{B}\right)\label{different}
\end{equation}
is the probability to get clicks on detectors in different channels.
Here $P_\mathrm{i_\mathrm{A},
i_\mathrm{B}}\!\left(\theta_\mathrm{A}, \theta_\mathrm{B}\right)$ is
the probability of registering clicks at the detectors
$i_\mathrm{A}=\{\mathrm{T_A},\mathrm{R_A}\}$ and
$i_\mathrm{B}=\{\mathrm{T_B},\mathrm{R_B}\}$ for the polarization
angles $\theta_\mathrm{A}$ and $\theta_\mathrm{B}$. According to the
photodetection theory, these probabilities are given by
\begin{equation}
P_{i_\mathrm{A}, i_\mathrm{B}}\left(\theta_\mathrm{A},
\theta_\mathrm{B}\right)=\sum\limits_{n,m=1}^{+\infty}\Tr\left(
\hat{\Pi}_{i_\mathrm{A}}^{(n)}\hat{\Pi}_{i_\mathrm{B}}^{(m)}
\hat{\Pi}_{j_\mathrm{A}}^{(0)}\hat{\Pi}_{j_\mathrm{B}}^{(0)}\hat\varrho\right),
\label{Probability}
\end{equation}
$i_\mathrm{A}\neq j_\mathrm{A}$, $i_\mathrm{B}\neq j_\mathrm{B}$,
where $\hat{\varrho}$ is the density operator,
\begin{equation}
\hat{\Pi}_{i_\mathrm{A(B)}}^{(n)}=:\frac{\left(\eta
\hat{n}_{i_\mathrm{A(B)}}+N\right)^n}{n!}
\exp\left[-\eta\hat{n}_{i_\mathrm{A(B)}}-N\right]:\label{POVM}
\end{equation}
is the positive operator-valued measure for the detector
$i_\mathrm{A(B)}$, $N$ is the mean number of stray-light and dark
counts~[Semenov et al., Phys. Rev. A {\bf 78}, 055803 (2008)],
$\eta$ is the detection efficiency, $:\cdots:$ means normal
ordering, and
\begin{equation}
\hat{n}_{i_\mathrm{A(B)}}
=\hat{a}^\dag_{i_\mathrm{A(B)}}\hat{a}_{i_\mathrm{A(B)}}.\label{PhotonNumber}
\end{equation}
are the photon-number operators at the outputs of the polarizing
beam-splitters $\mathrm{PBS}$. The operator input-output relations
of the polarization analyzers,
\begin{eqnarray}
\hat{a}_{\scriptscriptstyle
T_\mathrm{A(B)}}=\hat{a}_{\scriptscriptstyle
\mathrm{H_{A(B)}}}\cos\theta_\mathrm{A(B)}+
\hat{a}_{\scriptscriptstyle \mathrm{V_{A(B)}}}
\sin\theta_\mathrm{A(B)}\label{IOR1},\\
\hat{a}_{\scriptscriptstyle
R_\mathrm{A(B)}}=-\hat{a}_{\scriptscriptstyle
\mathrm{H_{A(B)}}}\sin\theta_\mathrm{A(B)}+
\hat{a}_{\scriptscriptstyle
\mathrm{V_{A(B)}}}\cos\theta_\mathrm{A(B)}\label{IOR2},
\end{eqnarray}
relate the operators $\hat{a}_{i_\mathrm{A(B)}}$ with the field
operators of the corresponding horizontal and vertical modes on the
inputs of the polarization analyzers, $\hat{a}_{\scriptscriptstyle
\mathrm{H_{A(B)}}}$ and $\hat{a}_{\scriptscriptstyle
\mathrm{V_{A(B)}}}$, respectively. Finally, we utilize the
input-output relation for the atmospheric channel, cf.
Eq.~(\ref{IOR_op}), and express the operators
$\hat{a}_{\scriptscriptstyle \mathrm{H_{A(B)}}}$ and
$\hat{a}_{\scriptscriptstyle \mathrm{V_{A(B)}}}$ via the
field-operators $\hat{a}^\mathrm{in}_{\scriptscriptstyle
\mathrm{H_{A(B)}}}$ and $\hat{a}^\mathrm{in}_{\scriptscriptstyle
\mathrm{V_{A(B)}}}$ at the transmitter,
\begin{eqnarray}
\hat{a}_{\scriptscriptstyle
\mathrm{H_{A(B)}}}=T\hat{a}^\mathrm{in}_{\scriptscriptstyle
\mathrm{H_{A(B)}}}+\sqrt{1-T^2}\hat{c}_{\scriptscriptstyle
\mathrm{H_{A(B)}}},\label{IOR_O_H2}\\
\hat{a}_{\scriptscriptstyle
\mathrm{V_{A(B)}}}=T\hat{a}^\mathrm{in}_{\scriptscriptstyle
\mathrm{V_{A(B)}}}+\sqrt{1-T^2}\hat{c}_{\scriptscriptstyle
\mathrm{V_{A(B)}}},\label{IOR_O_V2}
\end{eqnarray}
where the fluctuating transmission coefficient $T$ is equal for all
four modes due to negligible depolarization in the Earth atmosphere
and small time delay between entangled photons.

\begin{figure}[ht!]
\includegraphics[clip=,width=\linewidth]{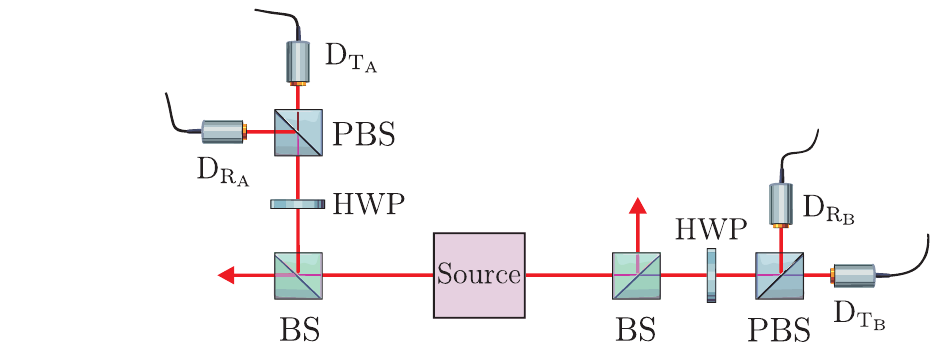}\\{\hfill\bf(a)}\\
\includegraphics[clip=,width=\linewidth]{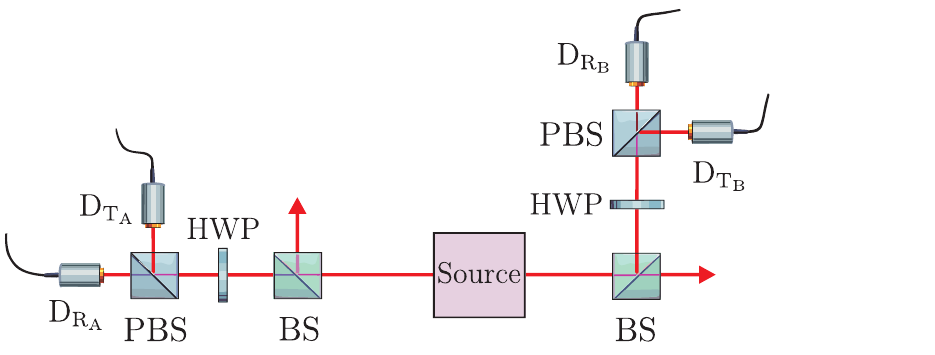}\\{\hfill\bf(b)}
\caption{\label{Fig:Bell_Replacement} Equivalent scenario for the
scheme presented in Fig.~\ref{Fig:Bell}. The time separation is
replaced by the spatial separation with two beam splitters
$\mathrm{BS}$ at each side. The schemes {\bf(a)} and {\bf(b)}
correspond to two kinds of events, which are detected by the
original device.}
\end{figure}

A special part of the device is the $50{:}50$ beam-splitter
$\mathrm{BS}$, cf.~Fig.~\ref{Fig:Bell}, which randomly separates
entangled photons. The devise registers only events, when the
photons left different output ports of the beam splitter. The other
half of the events, when the photons come to the same polarization
analyzer, are not resolved by this device. For simplicity, we can
consider an equivalent scenario, see
Fig.~\ref{Fig:Bell_Replacement}. Time-separated modes are replaced
by spatial-ones. At each side the modes are split by two $50{:}50$
beam splitters $\mathrm{BS}$. The resolved events in the original
device, cf.~Fig.~\ref{Fig:Bell}, correspond to the following events
in Fig.~\ref{Fig:Bell_Replacement}: (a) clicks at the reflection
port in the side $\mathrm{A}$ and the transmission port in the side
$\mathrm{B}$, (b) clicks at the transmission port in the side
$\mathrm{A}$ and the reflection port in the side $\mathrm{B}$. The
probability of both events is $2P_{i_\mathrm{A},
i_\mathrm{B}}\!\left(\theta_\mathrm{A}, \theta_\mathrm{B}\right)$,
which should be used instead of $P_{i_\mathrm{A},
i_\mathrm{B}}\!\left(\theta_\mathrm{A}, \theta_\mathrm{B}\right)$
with additional $50\%$ losses for the detection efficiency $\eta$
caused by the beam-splitters $\mathrm{BS}$. However, the factor $2$
disappears in Eq.~(\ref{correlation}). Hence, the main effect of the
beam-splitter $\mathrm{BS}$ consists in additional $50\%$ detection
losses.

The parametric down conversion source irradiates the states
\begin{equation}
\left|\mathrm{PDC}\right\rangle=(\cosh\chi)^{-2}\sum\limits_{n=0}^{+\infty}
\sqrt{n+1}\tanh^n\chi\left|\Phi_n\right\rangle,\label{PDC1}
\end{equation}
where $\chi$ is the squeezing parameter, and
\begin{eqnarray}
&&\left|\Phi_n\right\rangle=\label{PDC2}\\
&&\frac{1}{\sqrt{n+1}}\sum\limits_{m=0}^{n}(-1)^m
\left|n-m\right\rangle_\mathrm{H_A}\left|m\right\rangle_\mathrm{V_A}
\left|m\right\rangle_\mathrm{H_B}\left|n-m\right\rangle_\mathrm{V_B}\nonumber.
\end{eqnarray}
For $n=1$,
\begin{eqnarray}
\ket{\Phi_1}&=&\frac{1}{\sqrt{2}}\Big(\ket{1}_\mathrm{H_A}\ket{0}_\mathrm{V_A}
\ket{0}_\mathrm{H_B}\ket{1}_\mathrm{V_B}\Big.\label{BellState}\\
\Big.&-&\ket{0}_\mathrm{H_A}\ket{1}_\mathrm{V_A}
\ket{1}_\mathrm{H_B}\ket{0}_\mathrm{V_B}\Big)\nonumber\\
&\equiv&\frac{1}{\sqrt{2}}\Big(\ket{\mathrm{H}}_\mathrm{A}
\ket{\mathrm{V}}_\mathrm{B}-
\ket{\mathrm{V}}_\mathrm{A}\ket{\mathrm{H}}_\mathrm{B}
\Big),\nonumber
\end{eqnarray}
is the Bell state, which for $\big(\theta_\mathrm{A}^{(1)},
\theta_\mathrm{B}^{(1)},\theta_\mathrm{A}^{(2)},
\theta_\mathrm{B}^{(2)}\big)=
\big(0,\frac{\pi}{8},\frac{\pi}{4},\frac{3\pi}{8}\big)$ maximally
violates the Bell inequality. It is convenient to solve the problem
using the Glauber-Sudarshan $P$ representation. The corresponding
characteristic function for the state~(\ref{PDC1}) reads as
\begin{widetext}
\begin{eqnarray}
\Phi_\mathrm{in}\left(\beta_\mathrm{\scriptscriptstyle
H_{A}},\beta_\mathrm{\scriptscriptstyle V_{A}},
\beta_\mathrm{\scriptscriptstyle
H_{B}},\beta_\mathrm{\scriptscriptstyle
V_{B}}\right)&=&\exp\left[-\frac{\tanh^{2}\chi
\left|\beta_\mathrm{\scriptscriptstyle V_{A}}\right|^2+\tanh^{2}\chi
\left|\beta_\mathrm{\scriptscriptstyle
H_{B}}\right|^2-\tanh\chi\left(\beta_\mathrm{\scriptscriptstyle
V_{A}}\beta_\mathrm{\scriptscriptstyle H_{B}}+
\beta^\ast_\mathrm{\scriptscriptstyle
V_{A}}\beta^\ast_\mathrm{\scriptscriptstyle H_{B}}
\right)}{1-\tanh^{2}\chi}\right]\label{CharFuncPDC}\\
&\times& \exp\left[-\frac{\tanh^{2}\chi
\left|\beta_\mathrm{\scriptscriptstyle H_{A}}\right|^2+\tanh^{2}\chi
\left|\beta_\mathrm{\scriptscriptstyle
V_{B}}\right|^2+\tanh\chi\left(\beta_\mathrm{\scriptscriptstyle
H_{A}}\beta_\mathrm{\scriptscriptstyle V_{B}}+
\beta^\ast_\mathrm{\scriptscriptstyle
H_{A}}\beta^\ast_\mathrm{\scriptscriptstyle V_{B}}
\right)}{1-\tanh^{2}\chi}\right].\nonumber
\end{eqnarray}
Similar to Appendix~\ref{App:QSIOR}, we utilize
Eqs.~(\ref{IOR1})-(\ref{IOR_O_V2}) and quantum-state input-output
relations~(\ref{IOR_ChF}). This results in the characteristic
function at the outputs of the polarization analyzers,
\begin{eqnarray}
\Phi_\mathrm{out}\left(\beta_{\scriptscriptstyle
T_\mathrm{A}},\beta_{\scriptscriptstyle R_\mathrm{A}},
\beta_{\scriptscriptstyle T_\mathrm{B}},\beta_{\scriptscriptstyle
R_\mathrm{B}}\right)&=&\int\limits_{0}^{1}\D T
\mathcal{P}\!\left(T\right)\exp\left[-\frac{T^2\tanh^{2}\chi
\left(\left|\beta_{\scriptscriptstyle
T_\mathrm{A}}\right|^2+\left|\beta_{\scriptscriptstyle
R_\mathrm{A}}\right|^2+\left|\beta_{\scriptscriptstyle
T_\mathrm{B}}\right|^2+\left|\beta_{\scriptscriptstyle
R_\mathrm{B}}\right|^2\right)}
{1-\tanh^{2}\chi}\right]\label{CharFuncPDCOut}\\
&\times&
\exp\left[-\frac{T^2\tanh\chi\left(\beta_{\scriptscriptstyle
T_\mathrm{A}}\beta_{\scriptscriptstyle T_\mathrm{B}}+
\beta_{\scriptscriptstyle
T_\mathrm{A}}^\ast\beta_{\scriptscriptstyle T_\mathrm{B}}^\ast+
\beta_{\scriptscriptstyle R_\mathrm{A}}\beta_{\scriptscriptstyle
R_\mathrm{B}}+ \beta_{\scriptscriptstyle
R_\mathrm{A}}^\ast\beta_{\scriptscriptstyle R_\mathrm{B}}^\ast
\right)\sin\left(\theta_\mathrm{B}-\theta_\mathrm{A}\right)}
{1-\tanh^{2}\chi}\right]\nonumber\\
&\times&
\exp\left[-\frac{T^2\tanh\chi\left(\beta_{\scriptscriptstyle
T_\mathrm{A}}\beta_{\scriptscriptstyle R_\mathrm{B}}+
\beta_{\scriptscriptstyle
T_\mathrm{A}}^\ast\beta_{\scriptscriptstyle R_\mathrm{B}}^\ast-
\beta_{\scriptscriptstyle R_\mathrm{A}}\beta_{\scriptscriptstyle
T_\mathrm{B}}- \beta_{\scriptscriptstyle
R_\mathrm{A}}^\ast\beta_{\scriptscriptstyle T_\mathrm{B}}^\ast
\right)\cos\left(\theta_\mathrm{B}-\theta_\mathrm{A}\right)}
{1-\tanh^{2}\chi}\right]\nonumber.
\end{eqnarray}

In terms of the characteristic function, Eq.~(\ref{Probability}) is
given by%
\setlength\arraycolsep{0pt}
\begin{equation}
P_{i_\mathrm{A}, i_\mathrm{B}}\!\left(\theta_\mathrm{A},
\theta_\mathrm{B}\right)=\int\limits_{-\infty}^{+\infty}\D^8\beta\,
\Phi_\mathrm{out}\left(\beta_{\scriptscriptstyle
T_\mathrm{A}},\beta_{\scriptscriptstyle R_\mathrm{A}},
\beta_{\scriptscriptstyle T_\mathrm{B}},\beta_{\scriptscriptstyle
R_\mathrm{B}}\right)
K_\mathrm{C}\!\left(\beta_{ i_\mathrm{A}}\right)%
K_\mathrm{C}\!\left(\beta_{ i_\mathrm{B}}\right)%
K_0\!\left(\beta_{j_\mathrm{A}}\right)%
K_0\!\left(\beta_{j_\mathrm{B}}\right),%
\label{ProbabilityCharF}
\end{equation}
where $\D^8\beta=\D^2\beta_{\scriptscriptstyle T_\mathrm{A}}
\D^2\beta_{\scriptscriptstyle R_\mathrm{A}}
\D^2\beta_{\scriptscriptstyle T_\mathrm{B}}
\D^2\beta_{\scriptscriptstyle R_\mathrm{A}}$,
\begin{equation}
K_0\!\left(\beta\right)=\frac{1}{\pi\eta}
\exp\left[-\frac{\left|\beta\right|^2}{\eta}-N\right],\label{K0}
\end{equation}
\begin{equation}
K_\mathrm{C}\!\left(\beta\right)=\delta\!\left(\beta\right)-
K_0\!\left(\beta\right).\label{KC}
\end{equation}
Substituting (\ref{CharFuncPDCOut}) in (\ref{ProbabilityCharF}) we
obtain an explicit form for the probabilities of simultaneous clicks
at both sides,
\begin{equation}
P_\mathrm{i_\mathrm{A}, i_\mathrm{B}}\left(\theta_\mathrm{A},
\theta_\mathrm{B}\right)=\left(1-\tanh^2\chi\right)^4
\left[\left\langle\frac{\exp\left(-2N\right)}
{C_\mathrm{0}+2C_\mathrm{1}+C_\mathrm{i_\mathrm{A},i_\mathrm{B}}}
\right\rangle-
\left\langle\frac{2\exp\left(-3N\right)}{C_\mathrm{0}+C_\mathrm{1}}
\right\rangle
+\left\langle\frac{\exp\left(-4N\right)}{C_\mathrm{0}}\right\rangle\right],
\label{ProbabilitySpecial}
\end{equation}
 where $\left\langle\ldots\right\rangle$ means averaging with
 respect to $T$,
\begin{equation}
C_\mathrm{0}=\left\{\eta^2T^4\tanh^2\chi- \left[1+\left(\eta
T^2-1\right)\tanh^2\chi\right]^2\right\}^2,\label{C0}
\end{equation}
\begin{equation}
C_\mathrm{1}=\eta T^2\left(1-\eta
T^2\right)\left(1-\tanh^2\chi\right)\tanh^2\chi
\left\{\eta^2T^4\tanh^2\chi- \left[1+\left(\eta
T^2-1\right)\tanh^2\chi\right]^2\right\},\label{C1}
\end{equation}
\begin{equation}
C_\mathrm{T_A,T_B}=C_\mathrm{R_A,R_B}=\eta^2
T^4\tanh^2\chi\left[1-\tanh^2\chi\right] ^2\left[\left(1-\eta
T^2\right)^2 \tanh^2\chi-\sin^2
\left(\theta_\mathrm{A}-\theta_\mathrm{B}\right)\right],\label{Csame}
\end{equation}
\begin{equation}
C_\mathrm{T_A,R_B}=C_\mathrm{R_A,T_B}=\eta^2
T^4\tanh^2\chi\left[1-\tanh^2\chi\right]^2\left[\left(1-\eta
T^2\right)^2 \tanh^2\chi-\cos^2
\left(\theta_\mathrm{A}-\theta_\mathrm{B}\right)\right].\label{Cdifferent}
\end{equation}
Equation~(\ref{ProbabilitySpecial}) can now be used in
Eqs.~(\ref{correlation})-(\ref{different}), and then inserted in
expression for the Bell parameter~(\ref{BellParameter}). Plots for
the Bell parameter as a function of the squeezing parameter $\chi$
is given in Fig.~3.
\end{widetext}

From the paper~\cite{Fedrizzi} we can extract some values of
parameters. Detection efficiency is $\eta=0.25$ ($6\,\mathrm{dB}$).
Additionally we have to include the beam splitter $\mathrm{BS}$
losses. The total efficiency of the receiver module is $\eta=0.125$
($9\,\mathrm{dB}$). The dark-count rate is $200\,\mathrm{s}^{-1}$
and the stray-light rate is $200\,\mathrm{s}^{-1}$. Coincidences are
integrated over a $1.25\,\mathrm{ns}$ time window. This gives the
mean number of stray-light and dark counts of
$N{=}5{\times}10^{-7}$. The mean losses of the atmospheric channel
are $\left\langle T^2\right\rangle{=}6.3{\times}10^{-4}$
($32\,\mathrm{dB}$).

\section{Quadrature and photon-number squeezing}
\label{App:Sq}

In this Appendix we consider some mathematical details related to
the post-selection procedure, which improves quadrature and
photon-number squeezing. We will characterize the photon-number
squeezing by the Mandel parameter,
\begin{equation}
Q{=}\frac{\left\langle:\Delta\hat
n^2:\right\rangle}{\left\langle\hat
n\right\rangle},\label{MandelParDef}
\end{equation}
where $\hat{n}$ is the photon-number operator, $:\ldots:$ means
normal ordering. Similarly, the quadrature squeezing is
characterized by the normally-ordered variance
\begin{equation}
\left\langle:\Delta\hat{X}^2:\right\rangle\label{NormOrdQuadrVar}
\end{equation}
 of the quadrature
\begin{equation}
\hat X=\hat{a}+\hat{a}^\dag,\label{QuadratureDef}
\end{equation}
where $\hat{a}$ is the field annihilation operator. One can also use
the value of squeezing in $\mathrm{dB}$ relatively to the vacuum
noise,
\begin{equation}
-10\log_{10}\left(Q+1\right)
\end{equation}
for the photon-number squeezing, and
\begin{equation}
-10\log_{10}\left(\left\langle:\Delta\hat{X}^2:\right\rangle+1\right)
\end{equation}
for the quadrature squeezing.

Input-output relation~(\ref{IOmoments}) for the normally-ordered
moments can be applied for obtaining the input-output relations for
squeezing factors~(\ref{MandelParDef}) and (\ref{NormOrdQuadrVar}),
\begin{equation}
Q_\mathrm{out}{=}\frac{\langle T^4\rangle}{\langle T^2\rangle}
Q_\mathrm{in}+\frac{\langle\Delta \eta_{\scriptscriptstyle
T}^2\rangle}{\langle T^2\rangle} \langle\hat
n\rangle_\mathrm{in},\label{AppG_pnsq}
\end{equation}
\begin{equation}
\langle:\Delta\hat X^2:\rangle_\mathrm{out}{=} \langle T^2\rangle
\langle:\Delta\hat X^2:\rangle_\mathrm{in}+ \langle\Delta
T^2\rangle\langle \hat X\rangle_\mathrm{in}^2,\label{AppG_qsq}
\end{equation}
where $\langle\Delta T^2\rangle$ and $\langle\Delta
\eta_{\scriptscriptstyle T}^2\rangle$ are the variances of the
transmission coefficient $T$ and the transmission efficiency
$\eta_{\scriptscriptstyle T}=T^2$, respectively, cf. also
Refs.~\cite{Semenov2009, Semenov2011}. The first terms in
Eqs.~(\ref{AppG_pnsq}) and (\ref{AppG_qsq}) resemble the standard
attenuation. The second terms are non-negative and increase with
increasing $\langle \hat X_{\rm in}\rangle^2$ and $\langle\hat
n_{\rm in}\rangle$. They describe the effects of loss fluctuations.
Hence the nonclassicality diminishes by two effects: the
beam-positioning fluctuations and the brightness of the input light.

An appropriate post-selection by the transmission coefficient $T$
can improve the nonclassicality of detected fields. The
corresponding procedure excludes the events with $T\leq
T_\mathrm{min}$. The remaining events are described by the modified
PDTC,
\begin{equation}
\mathcal{P}_\mathrm{ps}\!\left(T\right)=%
\frac{1}{\overline{\mathcal F}}%
\begin{cases}
\mathcal{P}\!\left(T\right),&\textrm{for}\,\, T\in[ T_\mathrm{min},T_0]\\
0,&\textrm{else}
\end{cases},\label{PDTC_PS}
\end{equation}
where
\begin{equation}
\overline{\mathcal F}{=}\overline{\mathcal
F}\!\left(T_\mathrm{min}\right)\label{FTmin}
\end{equation}
is the value of exceedance~(\ref{ExceedanceCum}) at the point
$T{=}T_\mathrm{min}$. Post-selected PDTC~(\ref{PDTC_PS}) can be used
in Eqs.~(\ref{AppG_pnsq}) and (\ref{AppG_qsq}) for calculation of
the improved values of the corresponding squeezing factors.

In order to characterize the feasibility of the post-selection
procedure, we can explicitly express the squeezing
factors~(\ref{AppG_pnsq}) and (\ref{AppG_qsq}) [and post-selected
PDTC~(\ref{PDTC_PS})] via the exceedance $\overline{\mathcal F}$.
For this purpose we have to resolve Eq.~(\ref{FTmin}) with respect
to $ T_\mathrm{min}$ and substitute it in post-selected
PDTC~(\ref{PDTC_PS}). For the particular case of $d{=}0$ the
corresponding relation,
\begin{equation}
{T_\mathrm{min}}=
T_0\exp\left[-\frac{1}{2}\left(-\frac{2\sigma^2}{R^2}
\ln(1{-}\overline{\mathcal F})\right)^{\frac{\lambda}{2}}\right].
\end{equation}
is obtained explicitly.

\end{document}